# An eclipsing binary distance to the Large Magellanic Cloud accurate to 2 per cent

G. Pietrzyński[1,2], D. Graczyk[1], W. Gieren[1], I.B. Thompson[3], B., Pilecki[1,2], A. Udalski[2], I. Soszyński[2], S. Kozłowski[2], P. Konorski[2], K. Suchomska[2], G. Bono[4,5], P. G. Prada Moroni[6,7], S. Villanova[1], N. Nardetto[8], F. Bresolin[9], R.P. Kudritzki[9], J. Storm[10], A. Gallenne[1], R. Smolec[11], D. Minniti[12,13], M. Kubiak[2], M. Szymański[2], R. Poleski[2], Ł. Wyrzykowski[2], K. Ulaczyk[2], P. Pietrukowicz[2], M. Górski[2], P. Karczmarek[2]

1. *Universidad de Concepción, Departamento de Astronomìa, Casilla 160-C, Concepciòn, Chile*
2. *Warsaw University Observatory, Aleje Ujazdowskie 4, 00-478 Warszawa, Poland*
3. *Carnegie Observatories, 813 Santa Barbara Street, Pasadena, CA 91101-1292, USA*
4. *Dipartimento di Fisica Universita' di Roma Tor Vergata, via della Ricerca Scientifica 1, 00133 Rome, Italy*
5. *INAF-Osservatorio Astronomico di Roma, Via Frascati 33, 00040 Monte Porzio Catone, Italy*
6. *Dipartimento di Fisica Universita' di Pisa, Largo B. Pontecorvo 2, 56127 Pisa, Italy*
7. *INFN, Sez. Pisa, via E. Fermi 2, 56127 Pisa, Italy*
8. *Laboratoire Fizeau, UNS/OCA/CNRS UMR6525, Parc Valrose, 06108 Nice Cedex 2, France*
9. *Institute for Astronomy, 2680 Woodlawn Drive, Honolulu, HI 96822, USA*
10. *Leibniz Institute for Astrophysics, An der Sternwarte 16, 14482, Postdam, Germany*
11. *Nicolaus Copernicus Astronomical Centre, Bartycka 18, 00-716 Warszawa, Poland*
12. *Departamento de Astronomía y Astrofísica, Pontificia Universidad Católica de Chile, Vicuña Mackenna 4860, Casilla 306, Santiago 22, Chile*

13. *Vatican Observatory, V00120 Vatican City State, Italy*

**In the era of precision cosmology it is essential to determine the Hubble Constant with an accuracy of 3% or better[1,2]. Currently, its uncertainty is dominated by the uncertainty in the distance to the Large Magellanic Cloud (LMC) which as the second nearest galaxy serves as the best anchor point of the cosmic distance scale[2,3]. Observations of eclipsing binaries offer a unique opportunity to precisely and accurately measure stellar parameters and distances[4,5]. The eclipsing binary method was previously applied to the LMC[6,7] but the accuracy of the distance results was hampered by the need to model the bright, early-type systems used in these studies. Here, we present distance determinations to eight long-period, late-type eclipsing systems in the LMC composed of cool giant stars. For such systems we can accurately measure both the linear and angular sizes of their components and avoid the most important problems related to the hot early-type systems. Our LMC distance derived from these systems is demonstrably accurate to 2.2 % (49.97 ± 0.19 (statistical) ± 1.11 (systematic) kpc) providing a firm base for a 3 % determination of the Hubble Constant, with prospects for improvement to 2 % in the future.**

The modelling of early-type eclipsing binary systems consisting of hot stars is made difficult by the problem to obtain accurate flux calibrations for early-type stars, and by the degeneracy between the stellar effective temperatures and reddening[8,9]. As a result, the distances determined from such systems are of limited (~5-10%) accuracy. A better distance accuracy can be obtained using binary systems composed of cool stars; such systems among the frequent dwarf stars in the LMC are however too faint for an accurate analysis with present-day telescopes.

The OGLE team has been monitoring some 35 million stars in the field of the LMC for more than 16 years[10]. Based on this unique dataset we have detected a dozen extremely scarce very long period (60 – 772 days) eclipsing binary systems composed of intermediate-mass late-type giants located in a quiet evolutionary phase on the helium burning loop[11] (see Supplementary Table 1). These well detached systems provide an opportunity to use the full potential of eclipsing binaries as precise and accurate distance indicators, and to calibrate the zero point of the cosmic distance scale with an accuracy of about 2 % [5,12,13].

In order to achieve this goal, we observed 8 of these systems (see Figure 1) over the last 8 years, collecting high-resolution spectra with the MIKE echelle spectrograph at the 6.5-m Magellan Clay telescope at the Las Campanas Observatory, and with the HARPS spectrograph attached to the 3.6-m telescope of the European Southern Observatory on La Silla, together with near infrared photometry obtained with the 3.5-m New Technology Telescope located on La Silla.

The spectroscopic and OGLE V- and I-band photometric observations of the binary systems were then analysed using the 2007 version of the standard Wilson-Devinney (WD) code[14,15], in an identical manner as in our recent work on a similar system in the Small Magellanic Cloud[9]. Realistic errors to the derived parameters of our systems were obtained from extensive Monte Carlo simulations (see Figure 2). For all observed eclipsing binaries, their astrophysical parameters were determined with an accuracy of a few percent (see Supplementary Tables 2-9).

For late-type stars we can use the very accurately calibrated (2 %) relation between their surface brightness and V-K color to determine their angular sizes from optical (V) and near-infrared (K) photometry[16]. From this surface brightness-color relation (SBCR) we can derive angular sizes of the components of our binary systems directly from the definition of the surface brightness. Therefore the distance can be measured by combining the angular diameters of the binary components derived in this way with their corresponding linear dimensions obtained from the analysis of the spectroscopic and photometric data. The distances measured with this very simple but accurate one-step method are presented in Supplementary Table 12. The statistical errors of the distance determinations were calculated adding quadratically the uncertainties on absolute dimensions, V-K colors, reddening, and the adopted reddening law. The reddening uncertainty contributes very little (0.4 %) to the total error[17,11]. A significant change of the reddening law (from Rv = 3.1 to 2.7) causes an almost negligible contribution at the level of 0.3 %. The accuracy of the V-K color for all components of our eight binary systems is better than 0.014 mag (0.7 %). The resulting statistical errors in the distances are very close to 1.5 %, and are dominated by the uncertainty in the absolute dimensions. Calculating a weighted mean from the individual distances to the eight target eclipsing binary systems, we obtain a mean LMC distance of 49.88 ± 0.13 kpc.

Our distance measurement might be affected by the geometry and depth of the LMC. Fortunately, the geometry of the LMC is simple and well studied [18]. Since nearly all the eclipsing systems are located very close to the center of the LMC and to the line of nodes (see Fig. 3) we fitted the distance to the center of the LMC disk plane assuming its spatial orientation [18]. We obtained an LMC barycenter distance of 49.97 ± 0.19 kpc (see Figure 4), nearly identical to the simple weighted mean value, which shows that the geometrical structure of the LMC has no significant influence on our present distance determination.

The systematic uncertainty in our distance measurement comes from the calibration of the SBCR and the accuracy of the zero points in our photometry. The rms scatter on the current SBCR is 0.03 mag[13], which translates to a 2 % accuracy in the respective angular diameters of the component stars. Since the surface brightness depends only very weakly on metallicity[16,17], this effect contributes to the total error budget at the level of only 0.3 % [9]. Both optical (V) and near-infared (K) photometric zero points are accurate to 0.01 mag (0.5 %). Combining these contributions quadratically we determine a total systematic error of 2.1 % in our present LMC distance determination.

The LMC contains significant numbers of different stellar distance indicators, and being the second closest galaxy to our own offers us a unique opportunity to study these indicators with the utmost precision. For this reason this galaxy has an impressive record of several hundred distance measurements which have been carried out over the years[2,3,19]. Unfortunately, virtually all LMC distance determinations are dominated by systematic errors, with each method having its own sources of uncertainties. This prevents a calculation of the true LMC distance by simply taking the mean of the reported distances resulting from different techniques. Our present LMC distance measurement of 49.97 ± 0.19 (statistical) ± 1.11 (systematic) kpc (i.e. a true distance modulus of 18.493 mag ± 0.008(statistical) ± 0.047 (systematic)) agrees well, within the combined errors, with the most recent distance determinations to the LMC[19]. Our purely empirical method allows us to estimate both statistical and systematic errors in a very reliable way, which is normally not the case, particularly in distance determinations relying in part on theoretical predictions of stellar properties and their dependences on environment. In particular, our result provides a significant improvement over previous LMC distance determinations made using observations of eclipsing binaries[7,20]. These studies were based on early-type systems for which no empirical surface brightness-color relation is available, so they had to rely on theoretical models to determine the effective temperature. Our present determination is based on many (8) binary systems and does not resort to model predictions.

The classical approach to derive the Hubble constant consists in deriving an absolute calibration of the Cepheid Period-Luminosity Relation (CPLR) which is then used to determine the distances to nearby galaxies containing Type Ia supernovae (SNIa).[21] SNIa are excellent standard candles reaching out to the region of unperturbed Hubble flow once their peak brightnesses are calibrated this way, and provide the most accurate determination of $H_0$.[22] A yet alternative approach to calibrate the CPLR with Cepheids in the LMC is to calibrate it in our own Milky Way galaxy using Hubble Space Telescope (HST) parallax measurements of the nearest Cepheids to the Sun[23]. However, the resulting CPLR from that approach is less accurate for two reasons: first, the Cepheid sample with HST parallaxes is very small (ten stars) as compared to the Cepheid sample in the LMC (2000 stars), which can be used to establish the CPLR once the LMC distance is known. Second, the average accuracy of the HST Cepheid parallaxes is 8%[23] and suffers from systematics which are not completely understood, including Lutz-Kelker bias[24,25]. Therefore the currently preferred route to determine the Hubble constant is clearly the one using the very abundant LMC Cepheid population whose mean distance is now known, with the result of this work, to 2.2 %. This result reduces the uncertainty on $H_0$ to a very firmly established 3 %.

We have good reasons to believe that there is significant room to improve on our current 2.2% distance determination to the LMC by improving the calibration of the SBCR for late-type stars,[12,16] which is the dominant source of systematic error in our present determination. We are currently embarked on such a program, and a distance determination to the LMC accurate to 1% seems within reach once the SBCR calibration is refined, with its corresponding effect on improving the accuracy of $H_0$ even further. This is similar to the accuracy of the geometrical distance to the LMC which is to be delivered by GAIA in some 12 years from now. The eclipsing binary technique will then likely provide the best opportunity to check on the future GAIA measurements for possible systematic errors.

**Acknowledgements** We gratefully acknowledge financial support for this work from the BASAL Centro de Astrofisica y Tecnologias Afines (CATA), Polish Ministry of Science, the Foundation for Polish Science (FOCUS, TEAM), Polish National Science Centre, and the GEMINI-CONICYT fund. The OGLE project has received funding from the European Research Council "Advanced Grant" Program. It is a pleasure to thank the staff astronomers at Las Campanas and ESO La Silla who provided expert support in the data acquisition. We thank Jorge F. Gonzalez for making IRAF scripts rvbina and spbina available to us. We also thank O. Szewczyk and Z. Kołaczkowski for their help with some of the observations.

**Author Information**

Correspondence and requests for materials should be addressed to pietrzyn@astrouw.edu.pl

**Author Contributions** G.P. photometric and spectroscopic observations and reductions. D.G., spectroscopic observations, modelling, data analysis, W.G. Observations and data analysis, I.T. , observations, RV determination, data analysis, B.P. spectroscopic observations and reductions, RV measurements., A.U. , I.S and S. K. optical observations and data reductions. P.K., K.S., M.K., M.Sz., R.P., Ł., W., K.U.,

P.P., M.G., P.K. observations. G.B., P.G.P.M., N.N, F.B., R.P.K., J.S., A.G. and R.S. Data analysis. S.V., analysis of the spectra, G.P. and W.G. worked jointly to draft the manuscript with all authors reviewing and contributing to its final form

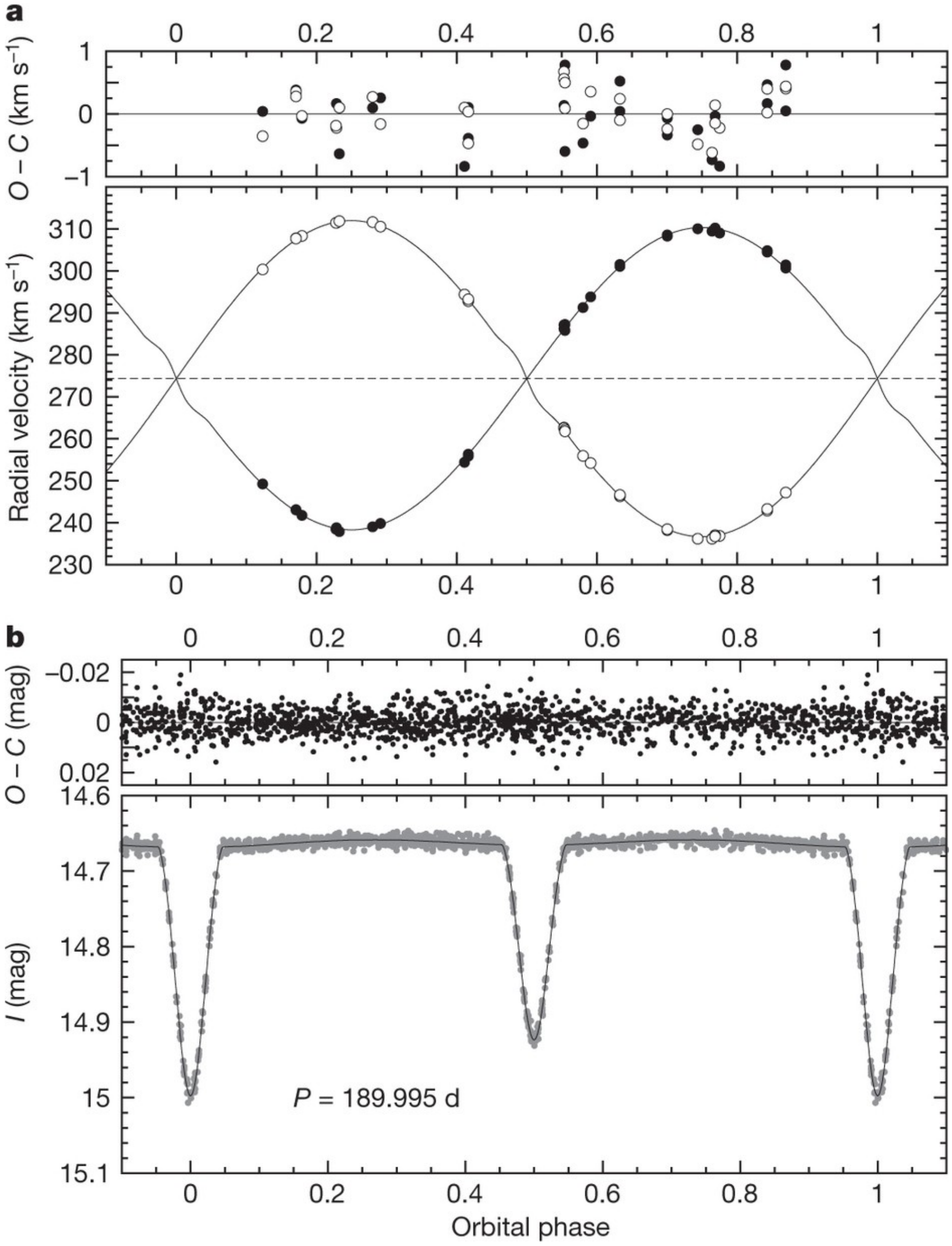

**Figure 1: Change of the brightness of the binary system OGLE-LMC-ECL-06575 and the orbital motion of its components.**

**a**, Main panel, orbital motion of the two binary components in the OGLE-LMC-ECL-06575 system. Filled and open circles, primary and secondary components, respectively. The top panel shows the residuals of the fit (see below): observed radial velocities (O) minus the computed radial velocities (C).

**b**, Main Panel, the I-band light curve (1200 epochs collected over 16 years) of the binary system OGLE-LMC-ECL-06575 together with the solution, as obtained with the Wilson-Devinney code. The top panel shows the residuals of the observed magnitudes from the computed orbital light curve.

All individual radial velocities were determined by the cross-correlation method using appropriate template spectra and the MIKE and HARPS spectra, yielding in all cases velocity accuracies better than 200 m/s (error bars smaller than the circles in the figure). The orbit (mass ratio, systemic velocity, velocity amplitudes, eccentricity, and periastron passage), was fitted with a least squares method to the measured velocities. The resulting parameters are presented in Supplementary Tables 2-9. The spectroscopic orbits, light curves and solutions for the remaining systems are of similar quality.

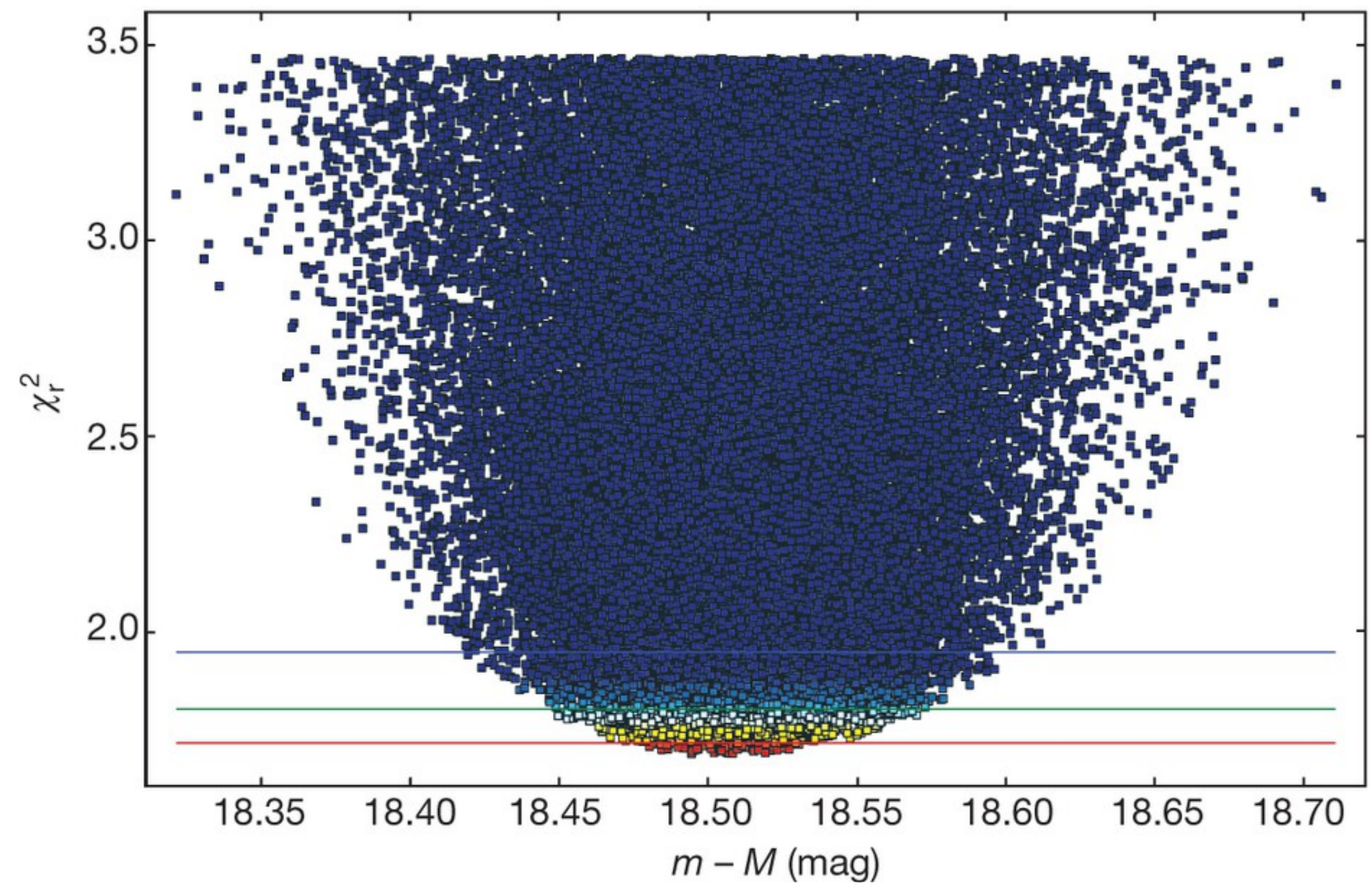

**Figure 2: Error estimation of the distance for one of our target binary systems.**

The reduced $\chi^2$ map for the OGLE-LMC-ECL-15260 system showing the dependence of the fit goodness to the V-band and I-band light curves on the distance modulus of the primary component. This map was obtained from 110,000 models computed with the Wilson-Devinney code[14,15] within a broad range of the radii $R_1$ and $R_2$, the orbital inclination $i$, the phase shift $\phi$, the secondary's temperature $T_2$ and the secondary's albedo $A_2$. In each case the distance $d$ was calculated from the V-band surface brightness – color (V–K) relation[16] and translated into distance modulus via formula $(m-M)=5\times\log(d)-5$. The horizontal lines correspond to the standard deviation limits of the derived distance modulus of 18.509 mag (50.33 kpc), accordingly from down to up: 1σ, 2σ and 3σ.

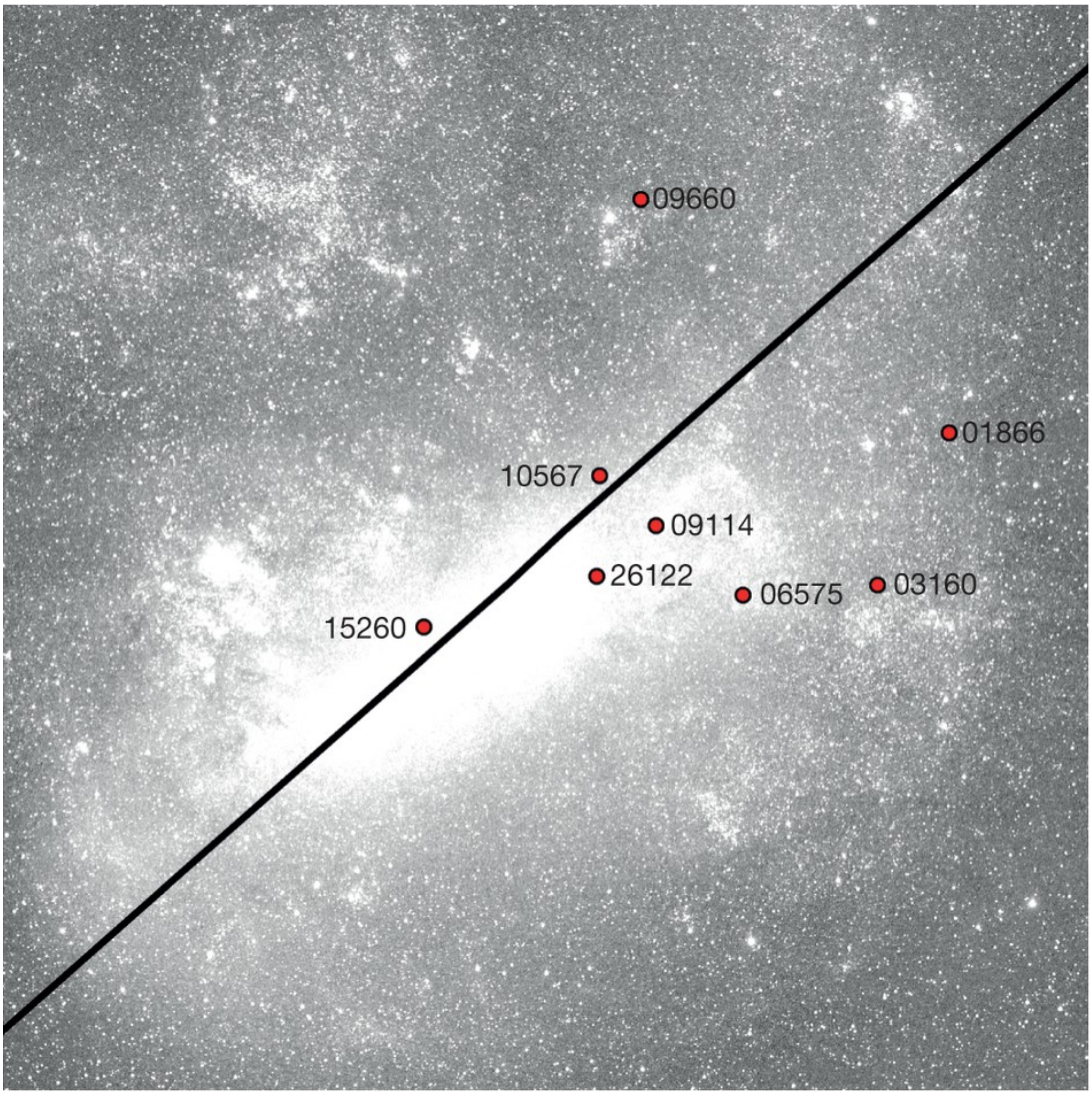

**Figure 3: Location of the observed eclipsing systems in the LMC.**

Most of our eight systems (marked as filled circles) are located quite close to the geometrical center of the LMC and to the line of nodes (marked with the line), resulting in very small corrections to the individual distances for the geometrical extension of this galaxy (in all cases smaller than the corresponding statistical error on the distance determination). The effect of the geometrical structure of the LMC on the mean LMC distance reported in our Letter is therefore negligible. The background image has a field of view of 8 x 8 degrees and is taken from the ASAS wide field sky survey[26].

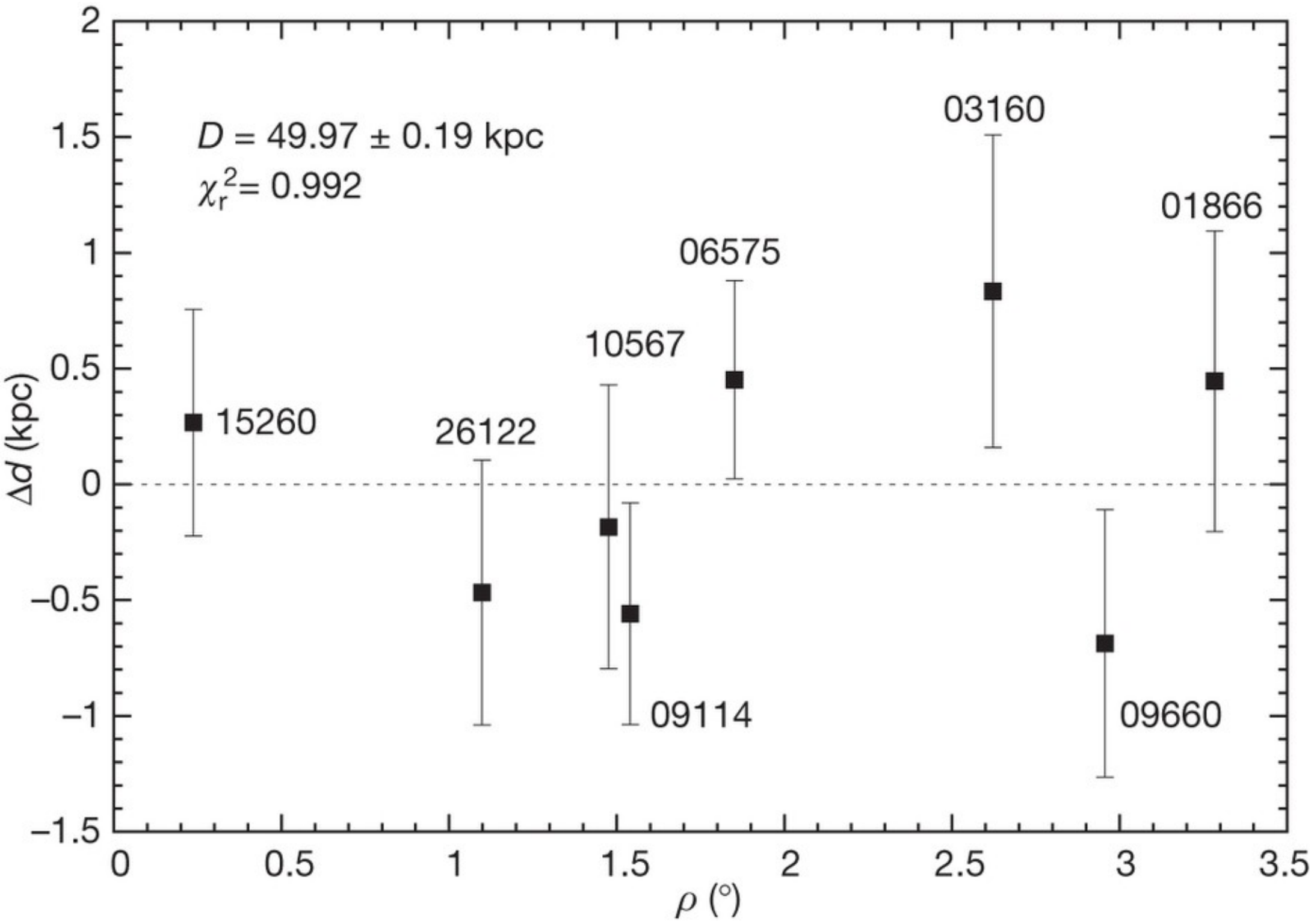

**Figure 4: Consistency among the distance determinations for the target binary systems**

Distance offsets between our particular eclipsing binary systems and the best fitted LMC disk plane, plotted against the angular distance of the systems from the LMC center. The identification of the systems is the same as in Figure 3. The error bars correspond to one sigma errors. We assumed the model of the LMC from van der Marel et al [18]. We fitted one parameter: the distance to the center of the LMC (R.A. = $5^h\ 25^m\ 06^s$, DEC = –69° 47' 00'') using a fixed spatial orientation of the LMC disk: inclination i = 28 deg and a position angle of the nodes of $\theta$ = 128 deg. The resulting distance to the LMC barycenter is 49.97 ± 0.19 kpc, with a reduced $\chi^2$ very close to unity.

## Supplementary Information:

### 1) Observations and target binary systems

All eclipsing systems studied in this letter, and listed in Supplementary Table 1, were discovered based on the OGLE-II and OGLE-III data[11]. Additional *V* and *I* band observations were collected with the Warsaw 1.3 m telescope at Las Campanas Observatory in the course of the OGLE IV project, and with the 1.3 m telescope at Cerro Tololo Observatory. Once the preliminary periods were calculated the new observations were secured mostly during eclipses, which resulted in a very good phase coverage of all targets. All photometric data were reduced with the image subtraction technique[29] and were calibrated based on the OGLE-III data[10]. The finding chart for all systems can be found on the OGLE Project webpage (ftp://ftp.astrouw.edu.pl/ogle/ogle3/OIII-CVS/lmc/ecl/fcharts)[11]. The raw OGLE V-band and I-band light curves are available at ftp://ftp.astrouw.edu.pl/ogle/ogle3/OIII-CVS/lmc/ecl/phot.

The near infrared photometry was performed with the ESO-La Silla 3.5 m NTT telescope equipped with the SOFI imager. Each system was observed outside of the eclipses during at least 5 different nights through *J* and *K* filters under photometric conditions together with a large number (12-16) of photometric standards from the UKIRT system[30]. The accuracy of the zero points obtained for every night was about 0.02 mag in both *J* and *K* bands. For more details regarding the observations, reduction and calibrations of the near-infrared data the reader is referred to (29).

High resolution echelle spectra were collected with the Las Campanas Observatory Magellan Clay 6.5 m telescope and the MIKE echelle spectrograph[32], and with the ESO 3.6 m telescope and the HARPS fiber-fed echelle spectrograph[33]. In the case of the MIKE observations, a 0.7 arcsec slit was used giving a resolution of about 40,000. The spectra were reduced with the dedicated pipeline software[34]. Exposure times ranged from 1200 sec to 3600 sec depending on observing conditions, and a typical resulting S/N ratio was between 7 and 30 at a wavelength of 450 nm. The HARPS observations were obtained at a resolution of 80,000 and a S/N of about 4-10 at 500 nm for integrations in the range from one minute to half an hour, and were reduced with the data reduction software developed by the Geneva observatory. Radial velocities were calculated with the Broadening Function (BF) formalism[35] and TODCOR[36]. We used templates from a library[37]. Templates were matched to the estimated mean effective temperature and gravity of the components. Both determinations agree with each other within 100 m/s.

| Id | R.A. (2000) | Dec. (2000) | P [days] |
|---|---|---|---|
| OGLE-LMC-ECL-09660 | $05^h\,11^m\,49^s.45$ | −67°05'45.2 | 167.6350 ± 0.0016 |
| OGLE-LMC-ECL-10567 | $05^h\,14^m\,01^s.89$ | −68°41'18.2 | 117.8708 ± 0.0012 |
| OGLE-LMC-ECL-26122 | $05^h14^m06^s.04$ | −69°15'56.9 | 771.7806 ± 0.0048 |
| OGLE-LMC-ECL-09114 | $05^h10^m19^s.64$ | −68°58'12.2 | 214.1707 ± 0.0009 |
| OGLE-LMC-ECL-06575 | $05^h04^m32^s.87$ | −69°20'51.0 | 189.8215 ± 0.0010 |
| OGLE-LMC-ECL-01866 | $04^h52^m15^s.28$ | −68°19'10.3 | 251.0068 ± 0.0043 |
| OGLE-LMC-ECL-03160 | $04^h55^m51^s.48$ | −68°13'48.0 | 150.0198 ± 0.0018 |
| OGLE-LMC-ECL-15260 | $05^h25^m25^s.66$ | −69°33'04.5 | 157.3243 ± 0.0008 |

**Supplementary Table 1. Selected eclipsing binary systems**.

The periods reported in this table and in all Supplementary Information are true orbital periods linked with the observed periods through the following formula:

$P = P_{OBS} \times \left(1 + \frac{\gamma}{c}\right)^{-1}$ where γ is the systemic velocity and c velocity of light, respectively.

## 2) The essentials of the modelling

The V and I-band light curves were cleaned from obvious outliers. No sigma clipping was applied. Both light curves and the two radial velocity curves – one per each component of a system – were analysed simultaneously for each system with the Wilson-Devinney code version 2007. We will denote by a subscript "1" the primary component and by a subscript "2" the secondary component of the system. We fitted the following set of adjustable parameters: the semi-major axis $a$, the mass ratio of the components $q$, the systemic velocity of the system γ, the apparent orbital period $P_{obs}$, the modified surface potentials $\Omega_1$ and $\Omega_2$, the secondary's mean surface temperature $T_2$, the internal luminosity of the primary component in the two bands $L1_V$, $L1_I$ and the orbital inclination $i$. In case of circular systems we fitted the epoch of the primary eclipse HJD0 and in some cases the albedo of the secondary $A_2$, while in the cases of eccentric systems we fitted the phase shift φ and additionally the orbital eccentricity $e$ and the argument of periastron ω. The initial values of adjustable parameters for the Wilson-Devinney code were found by a trial-and-error procedure.

The temperature of the primary $T_1$ (which is important because it scales the temperature of the secondary component and influences the limb darkening coefficients) was set as follows. The initial value was set to 5000 K, as this value is a reasonable assumption for late type giant stars. Then we run the WD code to obtain a preliminary model from which we obtained the components' light ratios in the *V*- and *I*-band. Additionally we extrapolated the model to calculate the *K*-band light ratio using theoretical limb darkening coefficients in this band. Combining these light ratios with the observed magnitudes in the *V*-, *I*- and *K*-bands, and using the reddenings derived from red clump stars in the fields containing our eclipsing binaries, we calculated the intrinsic colors $(V–I)_0$, $(V–K)_0$ of both stars. Using several calibrations between color indices and effective temperature[38,39,40,41] we estimated the temperatures of both components, especially the primary's temperature $T_1$. Then we run the WD code again with the new temperatures. Once updated reddening estimates were obtained we repeated the procedure. We iterated the temperature determination several times until full consistency of the model parameters was obtained, especially an agreement between the distance obtained from the surface brightness calibration and the distance resulting from bolometric flux scaling.

For most of our systems a correlation between the relative radii of both components can be observed. This is usual in the case of eclipsing binaries with partial eclipses. Two of our systems, LMC-ECL-09114 and LMC-ECL-09660, show total eclipses thus this correlation is unimportant in their case. However for the rest of the sample the influence of this correlation on the stellar radii and the resulting distance determination had to be investigated. To this end we first calculated approximate spectroscopic light ratios from the intensity of absorption lines and compared them to our light curve model predictions. A disagreement was found only for system LMC-ECL-10567. Subsequently we performed Monte Carlo simulations to investigate multi-dimensional parameter space and possible correlations between parameters. Some details of these simulations are given at the end of this section.

Initially we used a logarithmic limb darkening law in all cases utilizing pre-computed tables of theoretical limb darkening coefficients (setting LD= -2 in the WD2007 code). Additionally, we computed models with linear limb darkening (setting LD= -1) and also models where coefficients of the linear limb darkening law were treated as adjustable parameters (setting LD = +1). Usually models computed with linear and logarithmic law limb darkenings result in similar reduced $\chi^2$ values. However models with adjusted coefficients of linear limb darkening usually lead to considerably better fits to the light curves. Exceptions are the systems LMC-ECL-26122, LMC-ECL-01866 and LMC-ECL-10567. In the first one no improvement in the reduced $\chi^2$ can be seen. In the second case this procedure leads to a solution with very low coefficients (below zero) indicating limb brightening. In the last system the fitted coefficients are peculiar: unacceptably high and larger in the infrared than in optical. In those three cases we adopted the solution obtained with fixed coefficients of the logarithmic limb darkening law.

We tested the possible presence of a third light in our light curves and spectra. We could not detect any additional source of absorption lines in our spectra. We tested our spectra using the CCF, TODCOR and BF methods with different templates corresponding to a temperature range from 3500 to 7000 K but we failed to detect any third light source stronger that 1% of the combined signal of the two components of the systems. We also computed models setting the third light parameter *l*3 (being a fraction of the total observed flux) as an adjustable parameter of a solution for all our systems. Only in the case of LMC-ECL-10567 we found a small contribution of a third light in the optical *V*-band light curve, however in the *I*-band the third light contribution is insignificant. The third light in this system, if real, might be a faint blue companion or an optical blend ($V$~20 mag; $I$~21.5 mag) e.g. a white dwarf or even a QSO.

After inspection of the absorption lines broadening, which mostly comes from rotational broadening, we concluded that the rotation velocities of the components are consistent with both components being synchronized. Thus in all our computations the rotation parameter $F$ was fixed at 1.0 for both stars of a given eclipsing binary system, corresponding to the rotation being synchronous with the orbital period.

We estimated statistical errors on the parameters by performing Monte-Carlo simulations. We calculated a large number of models (usually over one hundred thousand per system) with input parameters randomly selected from a broad range of possible values and compared them to the *V*-band and *I*-band light curves to compute the residuals of the model and the resulting reduced $\chi^2$. In the Monte-Carlo simulations we allowed to vary the following parameters : the modified surface potentials $\Omega_1$ and $\Omega_2$, the secondary's mean surface temperature $T_2$, the orbital inclination i, the epoch of the primary eclipse HJD0 and in some cases, the albedo of the secondary $A_2$. The spectroscopic parameters during the simulations were kept fixed. The resulting $\chi^2$ maps and parameter correlation maps were used to derive realistic errors on the model parameters. For example, the radii correlation in the case of eclipsing binaries with partial eclipses is only marginally affecting our distances because the prime source of statistical error is the uncertainty of the sum of the radii $R_1$+$R_2$. Although for our systems, on average, radii are known with an accuracy of 3%, the sum of the radii is always known with an accuracy better than 2.0%. The largest source of uncertainty in the radii sum determination turns out to be the correlation between the orbital inclination and the radii.

| Parameter | Primary | Secondary |
|---|---|---|
| *(m-M)* = distance modulus | 18.489 | 18.489 |
| $M/M_\odot$ = mass | 2.969 ± 0.020 | 2.988 ± 0.018 |
| $R/R_\odot$ = *radius* | 23.75 ± 0.66 | 43.87 ± 1.14 |
| *T =effective temperature* | 5352 ± 70 K | 4677 ± 75 K |
| *K = velocity amplitude* | 35.13 ± 0.06 km/s | 34.91 ± 0.08 km/s |
| e = eccentricity | 0.0517 ± 0.0013 | |
| ω = periastron passage | 212.1 ± 1.5 deg | |
| γ = systemic velocity | 286.24 ± 0.04 km/s | |
| *P* =orbital period | 167.6350 ± 0.0016 days | |
| i =inclination | 87.81 ± 0.31 deg | |
| $a/R_\odot$ = orbit size | 232.00 ± 0.32 | |
| *q = mass ratio* | 1.0065 ± 0.0027 | |
| *A* = albedo | 0.5 (fixed) | |
| $x_V$ =linear limb darkening coeff. | 0.697 ± 0.098 | 0.726 ± 0.036 |
| $x_I$= linear limb darkening coeff. | 0.450 ± 0.077 | 0.391 ± 0.039 |
| *V* (observed magnitude) | 17.303 | 16.799 |
| *V−I* (observed color) | 0.959 | 1.234 |
| *V−K* (observed color) | 2.190 | 2.830 |
| *E(B-V)* reddening | 0.127 ± 0.020 | |
| [Fe/H] = metallicity | −0.44 ± 0.10 | |

**Supplementary Table 2. Astrophysical parameters of the OGLE-LMC-ECL-09660 system**

| Parameter | Primary | Secondary |
|---|---|---|
| *(m-M)* = distance modulus | 18.489 | 18.491 |
| $M/M_\odot$ =mass | 3.345 ± 0.040 | 3.183 ± 0.038 |
| $R/R_\odot$ = *radius* | 25.6 ± 1.6 | 36.0 ± 2.0 |
| *T =effective temperature* | 5067 ± 73 K | 4704 ± 80 K |
| *K = velocity amplitude* | 39.31 ± 0.13 km/s | 41.32 ± 0.14 km/s |
| e = eccentricity | 0.0 (fixed) | |
| ω = periastron passage | 90 (fixed) | |
| γ = systemic velocity | 265.10 ± 0.08 km/s | |
| *P* =orbital period | 117.8708 ± 0.0012 days | |
| i = inclination | 83.47 ± 0.33 deg | |
| $a/R_\odot$ = orbit size | 189.13 ± 0.45 | |
| *q = mass ratio* | 0.9515 ± 0.0043 | |
| *A* = albedo | 0.5 (fixed) | 0.100 ± 0.053 |
| $x_V$ =linear limb darkening coeff. | not adjusted | not adjusted |
| $x_I$= linear limb darkening coeff. | not adjusted | not adjusted |
| $l3_V$ = third light | 0.0482 ± 0.0305 | |
| $l3_I$ = third light | 0.0036 ± 0.0315 | |
| $l3_K$ = third light | 0.0 (fixed) | |
| *V* (observed magnitude) | 17.374 | 17.114 |
| *V−I* (observed color) | 1.019 | 1.158 |
| *V−K* (observed color) | 2.355 | 2.730 |
| *E(B-V)* reddening | 0.102 ± 0.020 | |

| | |
|---|---|
| [Fe/H] = metallicity | −0.81 ± 0.20 |

**Supplementary Table 3. Astrophysical parameters of the OGLE-LMC-ECL-10567 system**

| Parameter | Primary | Secondary |
|---|---|---|
| *(m-M)* = distance modulus | 18.470 | 18.468 |
| $M/M_{\odot}$ = mass | 3.593 ± 0.055 | 3.411 ± 0.047 |
| $R/R_{\odot}$ = *radius* | 32.71 ± 0.51 | 22.99 ± 0.48 |
| *T =effective temperature* | 4989 ± 80 K | 4995 ± 81 K |
| *K = velocity amplitude* | 23.80 ± 0.10 km/s | 25.08 ± 0.14 km/s |
| e = eccentricity | 0.4186 ± 0.0019 | |
| ω = periastron passage | 302.3 ± 0.2 deg | |
| γ = systemic velocity | 266.38 ± 0.07 km/s | |
| *P* =orbital period | 771.7806 ± 0.0048 days | |
| i = inclination | 88.45 ± 0.04 deg | |
| $a/R_{\odot}$ = orbit size | 677.64 ± 2.36 | |
| *q = mass ratio* | 0.9491 ± 0.0067 | |
| *A* = albedo | 0.5 (fixed) | |
| $x_V$ =linear limb darkening coeff. | not adjusted | not adjusted |
| $x_I$= linear limb darkening coeff. | not adjusted | not adjusted |
| *V* (observed magnitude) | 17.067 | 17.827 |
| $V{-}I$ (observed color) | 1.093 | 1.088 |
| $V{-}K$ (observed color) | 2.561 | 2.558 |
| $E(B\text{-}V)$ reddening | 0.140 ± 0.020 | |
| [Fe/H] = metallicity | −0.15 ± 0.10 | |

**Supplementary Table 4. Astrophysical parameters of the OGLE-LMC-ECL-26122 system**

| Parameter | Primary | Secondary |
|---|---|---|
| *(m-M)* = distance modulus | 18.465 | 18.465 |
| $M/M_{\odot}$ = mass | 3.303 ± 0.028 | 3.208 ± 0.026 |
| $R/R_{\odot}$ = *radius* | 26.18 ± 0.31 | 18.64 ± 0.30 |
| *T =effective temperature* | 5288 ± 81 K | 5470 ± 96 K |
| *K = velocity amplitude* | 32.76 ± 0.08 km/s | 33.37 ± 0.10 km/s |
| e = eccentricity | 0.0393 ± 0.0018 | |
| ω = periastron passage | 97.1 ± 0.3 deg | |
| γ = systemic velocity | 272.04 ± 0.05 km/s | |
| *P* =orbital period | 214.1707 ± 0.0009 days | |
| i =inclination | 88.77 ± 0.17 deg | |
| $a/R_{\odot}$ = orbit size | 281.38 ± 0.54 | |
| *q = mass ratio* | 0.9711 ± 0.0037 | |
| *A* = albedo | 0.5 (fixed) | |
| $x_V$ =linear limb darkening coeff. | 0.632 ± 0.068 | 0.555 ± 0.106 |
| $x_I$= linear limb darkening coeff. | 0.323 ± 0.067 | 0.319 ± 0.082 |
| $V$ (observed magnitude) | 17.217 | 17.783 |
| $V{-}I$ (observed color) | 1.059 | 0.984 |
| $V{-}K$ (observed color) | 2.319 | 2.188 |
| $E(B\text{-}V)$ reddening | 0.160 ± 0.020 | |
| [Fe/H] = metallicity | −0.23 ± 0.10 | |

**Supplementary Table 5. Astrophysical parameters of the OGLE-LMC-ECL-09114 system**.

| Parameter | Primary | Secondary |
|---|---|---|
| *(m-M)* = distance modulus | 18.497 | 18.497 |
| $M/M_{\odot}$ = mass | 4.152 ± 0.030 | 3.966 ± 0.032 |
| $R/R_{\odot}$ = *radius* | 39.79 ± 1.35 | 49.35 ± 1.45 |
| *T =effective temperature* | 4903 ± 72 K | 4681 ± 77 K |
| *K = velocity amplitude* | 36.03 ± 0.09 km/s | 37.72 ± 0.07 km/s |
| e = eccentricity | 0.0 (fixed) | |
| ω = periastron passage | 90 (fixed) | |
| γ = systemic velocity | 274.32 ± 0.05 km/s | |
| $P$ =orbital period | 189.8215 ± 0.0010 days | |
| i = inclination | 82.06 ± 0.13 deg | |
| $a/R_\odot$ = orbit size | 279.44 ± 0.44 | |
| *q = mass ratio* | 0.9552 ± 0.0034 | |
| *A* = albedo | 0.5 (fixed) | 0.648 ± 0.043 |
| $x_V$ =linear limb darkening coeff. | 0.789 ± 0.076 | 0.791 ± 0.068 |
| $x_I$= linear limb darkening coeff. | 0.336 ± 0.071 | 0.337 ± 0.065 |
| *V* (observed magnitude) | 16.642 | 16.490 |
| *V−I* (observed color) | 1.099 | 1.190 |
| *V−K* (observed color) | 2.529 | 2.775 |
| *E(B-V)* reddening | 0.107 ± 0.020 | |
| [Fe/H] = metallicity | −0.45 ± 0.10 | |

**Supplementary Table 6. Astrophysical parameters of the OGLE-LMC-ECL-06575 system**

| Parameter | Primary | Secondary |
|---|---|---|
| *(m-M)* = distance modulus | 18.496 | 18.496 |
| $M/M_\odot$ = mass | 3.575 ± 0.028 | 3.574 ± 0.038 |
| $R/R_\odot$ = *radius* | 28.20 ± 1.06 | 46.96 ± 0.61 |
| *T =effective temperature* | 5327 ± 72 K | 4541 ± 85 K |
| *K = velocity amplitude* | 33.27 ± 0.14 km/s | 33.28 ± 0.05 km/s |
| e = eccentricity | 0.2412 ± 0.0008 | |
| ω = periastron passage | 15.5 ± 0.3 | |
| γ = systemic velocity | 293.44 ± 0.04 km/s | |
| *P* =orbital period | 251.0068 ± 0.0043 days | |
| i =inclination | 83.34 ± 0.10 deg | |
| $a/R_\odot$ = orbit size | 322.68 ± 0.73 | |
| *q = mass ratio* | 0.9997 ± 0.0045 | |
| *A* = albedo | 0.5 (fixed) | |
| $x_V$ =linear limb darkening coeff. | not adjusted | not adjusted |
| $x_I$= linear limb darkening coeff. | not adjusted | not adjusted |
| *V* (observed magnitude) | 16.916 | 16.842 |
| *V−I* (observed color) | 0.952 | 1.250 |
| *V−K* (observed color) | 2.170 | 2.973 |
| *E(B-V)* reddening | 0.115 ± 0.020 | |
| [Fe/H] = metallicity | −0.70 ± 0.10 | |

**Supplementary Table 7. Astrophysical parameters of the OGLE-LMC-ECL-01866 system.**

| Parameter | Primary | Secondary |
|---|---|---|
| *(m-M)* = distance modulus | 18.505 | 18.505 |
| $M/M_\odot$ = mass | 1.792 ± 0.027 | 1.799 ± 0.028 |
| $R/R_\odot$ = *radius* | 16.36 ± 1.06 | 37.42 ± 0.52 |
| *T =effective temperature* | 4954 ± 83 K | 4490 ± 82 K |
| *K = velocity amplitude* | 30.47 ± 0.14 km/s | 30.35 ± 0.11 km/s |
| e = eccentricity | 0.0 (fixed) | |
| ω = periastron passage | 90 (fixed) | |
| γ = systemic velocity | 267.68 ± 0.08 km/s | |
| *P* =orbital period | 150.0198 ± 0.0018 days | |
| i = inclination | 83.36 ± 0.57 deg | |
| $a/R_{\odot}$ = orbit size | 182.01 ± 0.52 | |
| *q = mass ratio* | 1.0039 ± 0.0058 | |
| *A* = albedo | 0.5 (fixed) | |
| $x_V$ =linear limb darkening coeff. | 0.763 ± 0.174 | 0.796 ± 0.139 |
| $x_I$= linear limb darkening coeff. | 0.623 ± 0.112 | 0.567 ± 0.069 |
| *V* (observed magnitude) | 18.589 | 17.453 |
| *V−I* (observed color) | 1.076 | 1.308 |
| *V−K* (observed color) | 2.542 | 3.062 |
| *E(B-V)* reddening | 0.123 ± 0.020 | |
| [Fe/H] = metallicity | −0.48 ± 0.20 | |

**Supplementary Table 8. Astrophysical parameters of the OGLE-LMC-ECL-03160 system**.

| Parameter | Primary | Secondary |
|---|---|---|
| *(m-M)* = distance modulus | 18.509 | 18.509 |
| $M/M_\odot$ = mass | 1.440 ± 0.024 | 1.426 ± 0.022 |
| $R/R_\odot$ = *radius* | 23.51 ± 0.69 | 42.17 ± 0.33 |
| *T =effective temperature* | 4706 ± 87 K | 4320 ± 81 K |
| *K = velocity amplitude* | 27.67 ± 0.11 km/s | 27.93 ± 0.14 km/s |
| e = eccentricity | 0.0 (fixed) | |
| ω = periastron passage | 90 (fixed) | |
| γ = systemic velocity | 276.66 ± 0.06 km/s | |
| *P* =orbital period | 157.3243 ± 0.0008 days | |
| i =inclination | 82.99 ± 0.39 deg | |
| $a/R_\odot$ = orbit size | 174.25 ± 0.56 | |
| *q = mass ratio* | 0.9905 ± 0.0059 | |
| *A* = albedo | 0.5 (fixed) | 0.530 ± 0.031 |
| $x_V$ =linear limb darkening coeff. | 0.676 ± 0.092 | 0.936 ± 0.049 |
| $x_I$= linear limb darkening coeff. | 0.598 ± 0.060 | 0.660 ± 0.031 |
| *V* (observed magnitude) | 18.066 | 17.390 |
| *V−I* (observed color) | 1.149 | 1.392 |
| *V−K* (observed color) | 2.737 | 3.213 |
| *E(B-V)* reddening | 0.100 ± 0.020 | |
| [Fe/H] = metallicity | −0.47 ± 0.15 | |

**Supplementary Table 9. Astrophysical parameters of the OGLE-LMC-ECL-15260 system.**

### 3) Reddening determination and its influence on the distance determination

We used two different techniques to derive the reddening. For all our systems but one (LMC-ECL-15260) we managed to co-add and disentangle their spectra using the RaVeSpAn software developed by our group, implementing the method proposed in (42). The disentangled spectra were renormalized with the luminosity ratio of the secondary to the primary L2/L1(λ) from our best model. Based on the analysis of the spectra of individual components we calculated a standard set of atmospheric parameters (the effective temperature $T_{eff}$, gravity log g, microturbulance velocity $v_t$ and metallicity [Fe/H]). Synthetic *V–I* colors were calculated utilizing two calibrations between atmospheric parameters and the intrinsic color $(V-I)_0$ [40,41]. Comparing the colors obtained in this way to the corresponding observed colors, we calculated reddenings with a typical accuracy of 0.02 mag. In addition we used OGLE-III photometry[10] in order to calculate the mean red clump brightness in the *V* and *I* bands in 7x7 arcmin regions centered on our binary systems. In all cases, we were able to calculate the mean red clump magnitudes with an accuracy better than 0.01 mag. Following other studies of red clump stars[43] we assumed a *V-I* color of 0.72 (e.g. median color of RC stars in the LMC) as the color corresponding to the foreground reddening E(B-V) = 0.075 mag[44]. Then the reddening was determined from the observed color of red clump stars using the reddening law[45], which is appropriate for the LMC and was used in all calculations presented in our letter. In order to check our determinations we also calculated the reddening for three other fields where accurate reddenings were obtained based on Strömgren photometry[46], UBV photometry[8] and spectral analysis[22]. As can bee seen in Table 1, our determinations agree very well with the corresponding values of the reddenings reported in the literature. Finally, we estimated reddenings from the LMC reddening maps[47] based on OGLE-III photometry of RR Lyrae stars. All the determinations are in good agreement (see Table 2). As the final reddening we adopted the mean from the available determinations.

| Field | R.A. (2000) | Dec. (2000) | $E(B-V)_{RC}$ | $E(B-V)_{Lit}$ | Reference |
|---|---|---|---|---|---|
| P168.6 | $5^h29^m53^s$ | -69°09’23’’ | 0.151 | 0.140 | Larsen et al. 2000 |
| P169.3 | $5^h34^m48^s$ | -69°42’36’’ | 0.119 | 0.110 | Bonanos et al. 2011 |
| P126.4 | $5^h02^m40^s$ | -68°24’21’’ | 0.119 | 0.103 | Groenewegen & Salaris (2001) |

**Supplementary Table 10.** Comparison of the reddenings obtained with red clump stars to the corresponding values reported in the literature, for three fields in the LMC.

| System | $E(B-V)_{RC}$ | $E(B-V)_{RR\ Lyr}$ | $E(B-V)_{spec}$ | $E(B-V)_{adopted}$ |
|---|---|---|---|---|
| OGLE-LMC-ECL-09660 | 0.126 | 0.13 | 0.126 | 0.127 |
| OGLE-LMC-ECL-10567 | 0.126 | 0.106 | 0.075 | 0.102 |
| OGLE-LMC-ECL-26122 | 0.138 | 0.127 | 0.155 | 0.140 |
| OGLE-LMC-ECL-09114 | 0.139 | 0.131 | 0.21 | 0.160 |
| OGLE-LMC-ECL-06575 | 0.107 | 0.101 | 0.113 | 0.107 |
| OGLE-LMC-ECL-01866 | 0.119 | 0.15 | 0.075 | 0.115 |
| OGLE-LMC-ECL-03160 | 0.126 | 0.124 | 0.12 | 0.123 |
| OGLE-LMC-ECL-15260 | 0.11 | 0.091 | ---- | 0.100 |

**Supplementary Table 11.** Reddening determinations for our target eclipsing systems based on an analysis of their disentangled spectra, red clump stars, and RR Lyrae stars.

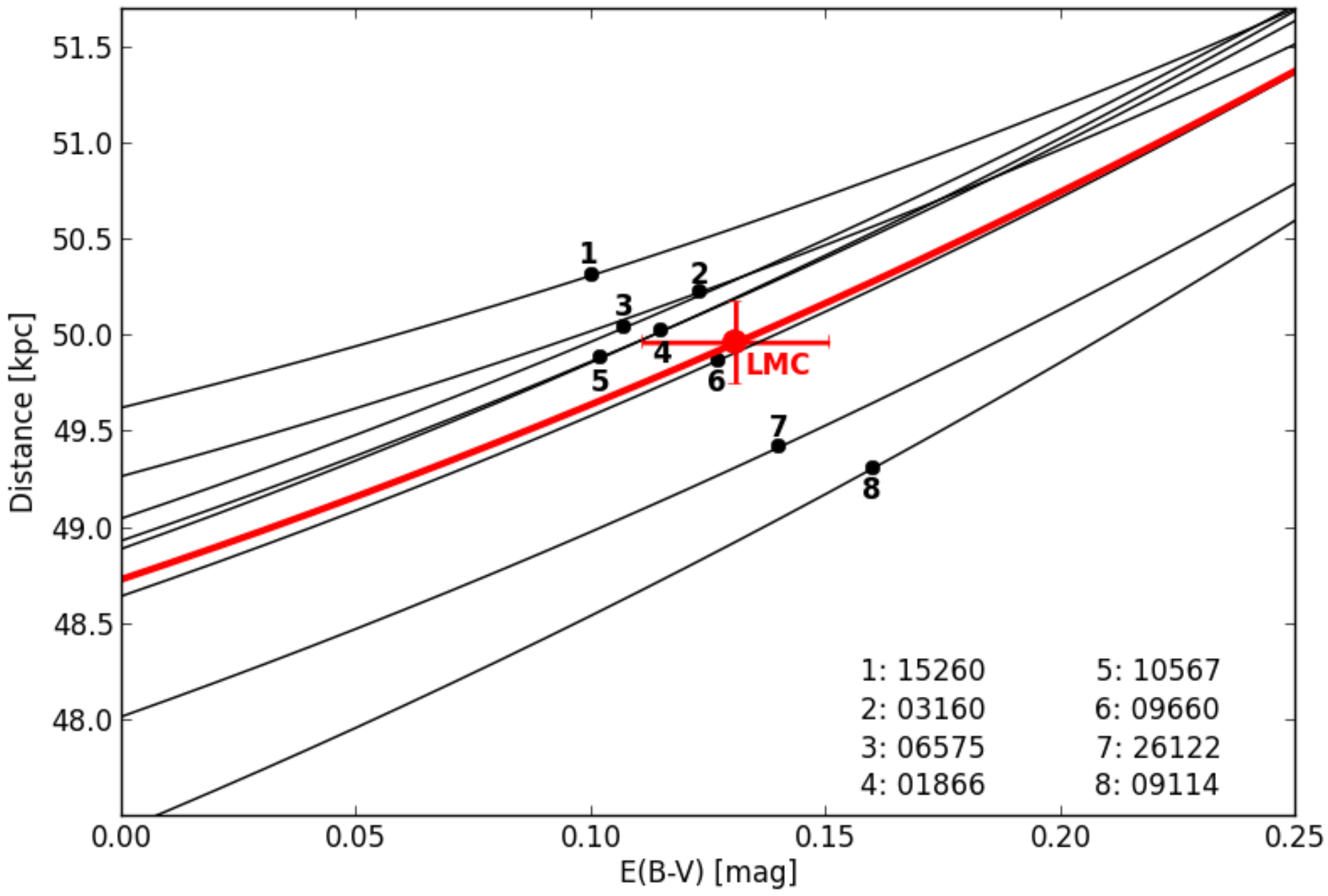

**Figure S1.** The final distances of the studied eclipsing systems calculated as a function of the adopted reddening. The thick red line is an average relation obtained from all systems. Points correspond to distances of individual systems identified in the plot legend. The larger red point denotes our distance determination to LMC. As can be appreciated from the error bars, a change of the reddening by 0.02 mag (1 σ) causes a change in the distance by 0.4% only (215 pc). This confirms the previous conclusions[9,17] that our method of distance determination is only very slightly dependent on the adopted reddening.

## 4) Distance determination

The surface brightnesses of the components of the studied systems were derived based on their disentangled (*V-K*) colors using the calibration $S_v = 2.656 + 1.483 \times (V-K)_0 - 0.044 \times (V-K)_0^2)$ [16]. The angular diameters were obtained directly from the definition of the surface brightness ( $\phi[mas] = 10^{0.2\times(S_v - V_0)}$ ). Finally distances to individual stars were derived combining their angular diameters obtained this way, and their linear diameters determined from the analysis of the spectroscopic and photometric data ( $d[pc] = 9.2984 \times \frac{R[R_{sun}]}{\phi[mas]}$ ).

In case of models with fixed theoretical limb darkening coefficients the disentangling of colors was performed in the way described in the Section 2. As a final distance we adopted the mean distance of both components of a system. The difference in the distance moduli of the components of the same system is not larger than 0.002 mag in all cases. Such small differences serve as an independent check of consistency and reliability of our results.

For models with adjusted limb darkening coefficients we employed another strategy to obtain the individual colors because we cannot extrapolate model predictions for the K-band in this case (there is no possibility to obtain appropriate limb darkening coefficients by their adjustment because of the lack of the K-band light curve). We assumed that both components of a system are at the same distance from us and we made iterations until we found a K-band components' light ratio fulfilling our assumption. Then we computed the disentangled (*V–K*) colors of the components.

In Supplementary Table 12 we list the individual and mean distances measured for all studied systems. As can be seen, the obtained distances are in excellent agreement.

| System | $(m\text{-}M)_1$ | $(m\text{-}M)_2$ | $(m\text{-}M)_{mean}$ | $\sigma_{m\text{-}M}$ | $\Delta(m\text{-}M)$ |
|---|---|---|---|---|---|
| OGLE-LMC-ECL-09660 | 18.489 | 18.489 | 18.489 | 0.025 | 0.035 |
| OGLE-LMC-ECL-10567 | 18.489 | 18.491 | 18.490 | 0.027 | 0.007 |
| OGLE-LMC-ECL-26122 | 18.470 | 18.468 | 18.469 | 0.025 | -0.005 |
| OGLE-LMC-ECL-09114 | 18.465 | 18.465 | 18.465 | 0.021 | -0.004 |
| OGLE-LMC-ECL-06595 | 18.497 | 18.497 | 18.497 | 0.019 | -0.021 |
| OGLE-LMC-ECL-01866 | 18.496 | 18.496 | 18.496 | 0.028 | -0.021 |
| OGLE-LMC-ECL-03160 | 18.505 | 18.505 | 18.505 | 0.029 | -0.012 |
| OGLE-LMC-ECL-15260 | 18.509 | 18.509 | 18.509 | 0.021 | 0.004 |

**Supplementary Table 12.** Individual distance moduli of the studied eclipsing binary systems. The symbols 1,2 and mean refer to the individual primary (1) and secondary (2) components of our eclipsing systems. The fifth column gives the total statistical uncertainty for the mean distance modulus. The geometrical corrections calculated from the model[18] are given in the last column.

| System | $(m\text{-}M)_{fix}$ | $(m\text{-}M)_{fit}$ | $\Delta(m\text{-}M)$ |
|---|---|---|---|
| OGLE-LMC-ECL-09660 | 18.490 | 18.489 | -0.001 |
| OGLE-LMC-ECL-10567 | 18.490 | 18.506 | 0.016 |
| OGLE-LMC-ECL-26122 | 18.469 | 18.482 | 0.013 |
| OGLE-LMC-ECL-09114 | 18.481 | 18.465 | -0.016 |
| OGLE-LMC-ECL-06575 | 18.52 | 18.497 | -0.023 |
| OGLE-LMC-ECL-01866 | 18.496 | 18.496 | 0.000 |
| OGLE-LMC-ECL-03160 | 18.511 | 18.505 | -0.006 |
| OGLE-LMC-ECL-15260 | 18.500 | 18.509 | 0.009 |

**Supplementary Table 13.** Comparison of the distance moduli of our systems as determined by fitting limb darkening coefficients (fit), and by using theoretical values of the limb darkening coefficients (fix). The average of the differences listed in the third column is -0.00075, which clearly shows that the way the limb darkening is chosen for our systems has a negligible effect on the final distance determination presented in our letter.